\documentclass[sigconf]{acmart}

\usepackage{xcolor,colortbl}

\definecolor{Gray}{gray}{0.85}
\definecolor{LightCyan}{rgb}{0.88,1,1}
\newcolumntype{a}{>{\columncolor{Gray}}c}
\newcolumntype{b}{>{\columncolor{white}}c}

\usepackage{amssymb}
\usepackage{hyperref}
\usepackage{epstopdf}

\usepackage{graphicx}
\usepackage[caption=false]{subfig}

\usepackage{enumerate}
\usepackage{slashbox}
\usepackage{amsmath}
\usepackage{verbatim}

\usepackage{microtype}
\DisableLigatures{encoding = *, family = * }

\usepackage[nolist]{acronym}
\begin{acronym}
	\acro{AAA}{Authentication, Authorization and Accounting}
	\acro{ACL}{Access Control List}
	\acro{AKI}{Accountable Key Infrastructure}
	\acro{API}{Application Programming Interface}
    \acro{BSM}{Basic Safety Message}
    \acro{BYOD}{Bring Your Own Device}
    \acro{BF}{Bloom Filter}
	\acro{C2C-CC}{Car2Car Communication Consortium}
	\acro{C2I}{Car-to-Infrastructure}
	\acro{C$^2$RL}{Compressed \ac{CRL}}
	\acro{CA}{Certification Authority}
	\acro{CN}{Common Name}
	\acro{CAM}{Cooperative Awareness Message}
	\acro{CAMP VSC3}{Crash Avoidance Metrics Partnership Vehicle Safety Consortium}
	\acro{CIA}{Confidentiality, Integrity and Availability}
	\acro{CRL}{Certificate Revocation List}
	\acro{CDN}{Content Delivery Network}
	\acro{COCA}{Cornell OnLine Certification Authority}
	\acro{CSR}{Certificate Signing Request}
	\acro{DAA}{Direct Anonymous Attestation}
	\acro{DDoS}{Distributed DoS}
	\acro{DDH}{Decisional Diffie-Helman}
	\acro{DENM}{Decentralized Environmental Notification Message}
	\acro{DHT}{Distributed Hash Table}
	\acro{DL/ECIES}{Discrete Logarithm and Elliptic Curve Integrated Encryption Scheme}
	\acro{DoT}{Department of Transportation}
	\acro{DoS}{Denial of Service}
	\acro{DoT}{Department of Transportation}
	\acro{DPA}{Data Protection Agency}
	\acro{DSRC}{Dedicated Short Range Communication}
	\acro{DSS}{Digital Signature Standard}
	\acro{DTLS}{Datagram \ac{TLS}}
	\acro{ECU}{Electronic Control Unit}
	\acro{EDR}{Event Data Recorder}
	\acro{ETSI}{European Telecommunications Standards Institute}
	\acro{ECDSA}{Elliptic Curve Digital Signature Algorithm}
	\acro{ECC}{Elliptic Curve Cryptography}
	\acro{EVITA}{E-safety Vehicle Intrusion protected Applications}
	\acro{FOT}{Field Operational Testing}
	\acro{FPGA}{Field-Programmable Gate Array}
	\acro{GCP}{Google Cloud Platform}
	\acro{GKE}{Google Kubernetes Engine}
	\acro{GPA}{Global Passive Adversary}
	\acro{GN}{GeoNetworking}
	\acro{GS-VLR}{Group Signatures with Verifier Local Revocation}
	\acro{GS}{Group Signatures}
	\acro{GM}{Group Manager}
	\acro{GBA}{Generic Bootstrapping Architecture}
	\acro{GUI}{Graphic User Interface}
	\acro{HSM}{Hardware Security Module}
	\acro{HTTP}{Hypertext Transfer Protocol}
	\acro{IEEE}{Institute of Electrical and Electronics Engineers}
	\acro{IETF}{Internet Engineering Task Force}
	\acro{IoT}{Internet of Things}
	\acro{ITS}{Intelligent Transport System}
	\acro{IT}{Information Technologies}
	\acro{IMSI}{International Mobile Subscriber Identity}
	\acro{IMEI}{International Mobile Station Equipment Identity}
	\acro{IdP}{Identity Provider}
	\acro{IDS}{Intrusion Detection System}
	\acro{ISP}{Internet Service Provider}
	\acro{LEA}{Law Enforcement Agency}
	\acro{LCPP}{Lightweight Conditional Privacy Preservation}
	\acro{LTC}{Long Term Certificate}
	\acro{LTCA}{Long Term CA}
	\acro{H-LTCA}{Home-LTCA}
	\acro{F-LTCA}{Foreign-LTCA}
	\acro{LDAP}{Lightweight Directory Access Protocol}
	\acro{LBS}{Location Based Service}
	\acro{LTE}{Long Term Evolution}
	\acro{LuST}{Luxembourg SUMO Traffic}
	\acro{MAC}{Message Authentication Code}
	\acro{MCA}{Message \ac{CA}}
	\acro{MEA}{Misbehavior Evaluation Authority}
	\acro{MANET}{Mobile Ad-hoc Network}
	\acro{MPB}{Most Pieces Broadcast}
	\acro{NoW}{Network on Wheel}
	\acro{OBU}{On-Board Unit}
	\acro{OEM}{Original Equipment Manufacturer}
	\acro{OCSP}{Online Certificate Status Protocol}
	\acro{PCA}{Pseudonym CA}
	\acro{PDP}{Policy Decision Point}
	\acro{PEP}{Policy Enforcement Point}
	\acro{PIR}{Private Information Retrieval}
	\acro{PKC}{Public Key Cryptography}
	\acro{PKCS}{Public Key Cryptosystem}
	\acro{PKI}{Public-Key Infrastructure}
	\acro{PRECIOSA}{Privacy Enabled Capability in Co-operative Systems and Safety Applications}
	\acro{PRESERVE}{Preparing Secure Vehicle-to-X Communication Systems}
	\acro{P2P}{peer-to-peer}
	\acro{PS}{Participatory Sensing}
	\acro{RA}{Resolution Authority}
	\acro{REST}{Representational State Transfer}
	\acro{RBAC}{Role Based Access Control}
	\acro{RCA}{Root \acs{CA}}
	\acro{RSU}{Roadside Unit}
	\acro{SAML}{Security Assertion Markup Language}
	\acro{SAS}{Sample Aggregation Service}
	\acro{SCMS}{Security Credential Management System}
	\acro{SCORE@F}{Système COopératif Routier Expérimental Français}
	\acro{SDSI}{Simple Distributed Security Infrastructure}
	\acro{SLA}{Service Level Agreement}
	\acro{SRAAC}{Secure Revocable Anonymous Authenticated Inter-Vehicle Communication}
	\acro{SeVeCom}{Secure Vehicle Communication}
	\acro{SIT}{Sichere Informationstechnologie}
	\acro{SLC}{Short-Lived Certificate}
	\acro{SoA}{Service-oriented-Approach}
	\acro{SIFS}{Short Inter Frame Space}
	\acro{SSO}{Single-Sign-On}
	\acro{SSL}{Secure Sockets Layer}
	\acro{SOAP}{Simple Object Access Protocol}
	\acro{TACK}{Temporary Anonymous Certified Key}
	\acro{TS}{Task Service}
	\acro{TLS}{Transport Layer Security}
	\acro{TPM}{Trusted Platform Module}
	\acro{TTP}{Trusted Third Party}
	\acro{TVR}{Ticket Validation Repository}
	\acro{URI}{Uniform Resource Identifier}
	\acro{VANET}{Vehicular Ad-hoc Network}
	\acro{V2I}{Vehicle-to-Infrastructure}
	\acro{V2V}{Vehicle-to-Vehicle}
	\acro{V2X}{\ac{V2V}/\ac{V2I}}
	\acro{VC}{Vehicular Communication}
	\acro{VM}{Virtual Machine}
	\acro{VSS}{\ac{VC} Security Subsystem}
	\acro{WAVE}{Wireless Access in Vehicular Environments}
	\acro{WSDL}{Web Services Discovery Language}
	\acro{W3C}{World Wide Web Consortium}
	\acro{V}{Vehicle}
	\acro{VANET}{Vehicular Ad-hoc Network}
	\acro{VLR}{Verifier-Local Revocation}
	\acro{VPKI}{Vehicular Public-Key Infrastructure}
	\acro{VPKIaaS}{VPKI as a Service}
	\acro{VM}{Virtual Machine}
	\acro{WS}{Web Service}
	\acro{WoT}{Web of Trust}
	\acro{WSACA}{\ac{WAVE} Service Advertisement \ac{CA}}
	\acro{XML}{Extensible Markup Language}
	\acro{XACML}{eXtensible Access Control Markup Language}
	\acro{3G}{3rd Generation}
\end{acronym}

\usepackage{makecell}

\usepackage{arydshln}

\usepackage{xcolor}

\usepackage{algpseudocode}

\usepackage{setspace}

\errorcontextlines\maxdimen

\makeatletter
\newcommand*{\algrule}[1][\algorithmicindent]{\makebox[#1][l]{\hspace*{.5em}\thealgruleextra\vrule height \thealgruleheight depth \thealgruledepth}}%
\newcommand*{\thealgruleextra}{}
\newcommand*{\thealgruleheight}{.75\baselineskip}
\newcommand*{\thealgruledepth}{.25\baselineskip}

\newcount\ALG@printindent@tempcnta
\def\ALG@printindent{%
	\ifnum \theALG@nested>0
	\ifx\ALG@text\ALG@x@notext
	\else
	\unskip
	\addvspace{-1pt}
	\ALG@printindent@tempcnta=1
	\loop
	\algrule[\csname ALG@ind@\the\ALG@printindent@tempcnta\endcsname]%
	\advance \ALG@printindent@tempcnta 1
	\ifnum \ALG@printindent@tempcnta<\numexpr\theALG@nested+1\relax
	\repeat
	\fi
	\fi
}%
\usepackage{etoolbox}
\patchcmd{\ALG@doentity}{\noindent\hskip\ALG@tlm}{\ALG@printindent}{}{\errmessage{failed to patch}}
\makeatother

\newbox\statebox
\newcommand{\myState}[1]{%
	\setbox\statebox=\vbox{#1}%
	\edef\thealgruleheight{\dimexpr \the\ht\statebox+1pt\relax}%
	\edef\thealgruledepth{\dimexpr \the\dp\statebox+1pt\relax}%
	\ifdim\thealgruleheight<.75\baselineskip
	\def\thealgruleheight{\dimexpr .75\baselineskip+1pt\relax}%
	\fi
	\ifdim\thealgruledepth<.25\baselineskip
	\def\thealgruledepth{\dimexpr .25\baselineskip+1pt\relax}%
	\fi
	\State #1%
	\def\thealgruleheight{\dimexpr .75\baselineskip+1pt\relax}%
	\def\thealgruledepth{\dimexpr .25\baselineskip+1pt\relax}%
}

\usepackage{footnote}

\usepackage{rotating}

\usepackage{soul}

\usepackage{algorithm}

\usepackage{tikz}
\usetikzlibrary{shadows,positioning,calc}
\tikzset{multiple/.style = {double copy shadow={shadow xshift=1ex,shadow
			yshift=-1.5ex,draw=black!30},fill=white,draw=black,thick,minimum height = 1cm,minimum
		width=2cm},
	ordinary/.style = {rectangle,draw,thick,minimum height = 1cm,minimum width=2cm}}

\usetikzlibrary{decorations.pathreplacing}

\usepackage{booktabs}
\usepackage{multirow}
\usepackage{siunitx}

\usepackage{algpseudocode}

\makeatletter
\renewcommand{\ALG@beginalgorithmic}{\footnotesize} 
\makeatother

\usepackage{afterpage}
\newlength{\oldtextfloatsep}

\renewcommand\footnotetextcopyrightpermission[1]{} 
\usepackage[export]{adjustbox}

\usepackage{booktabs} 

\setcopyright{rightsretained}

\begin{document}

\hyphenation{ano-nymi-ty pseu-do-nyms allo-ca-ting de-allo-ca-ting}

\title[DEMO: VPKIaaS: A Highly-Available and Dynamically-Scalable \\ \acl{VPKI}]{DEMO: VPKIaaS: A Highly-Available and Dynamically-Scalable \acl{VPKI}}


\author{Hamid Noroozi}
\orcid{}
\affiliation{%
	\institution{Networked Systems Security Group}
	\streetaddress{KTH Royal Institute of Technology}
	\city{KTH, Stockholm} 
	\state{Sweden} 
}
\email{hnoroozi@kth.se}

\author{Mohammad Khodaei}
\orcid{}
\affiliation{%
	\institution{Networked Systems Security Group}
	\streetaddress{KTH Royal Institute of Technology}
	\city{KTH, Stockholm} 
	\state{Sweden} 
}
\email{khodaei@kth.se}

\author{Panos Papadimitratos}
\affiliation{%
	\institution{Networked Systems Security Group}
	\streetaddress{KTH Royal Institute of Technology}
	\city{KTH, Stockholm} 
	\state{Sweden} 
}
\email{papadim@kth.se}


\begin{abstract}

The central building block of secure and privacy-preserving \ac{VC} systems is a \ac{VPKI}, which provides vehicles with multiple anonymized credentials, termed \emph{pseudonyms}. These pseudonyms are used to ensure message authenticity and integrity while preserving vehicle (and thus passenger) privacy. In the light of emerging large-scale multi-domain \ac{VC} environments, the efficiency of the \ac{VPKI} and, more broadly, its scalability are paramount. In this extended abstract, we leverage the state-of-the-art \ac{VPKI} system and \emph{enhance} its functionality towards a highly-available and dynamically-scalable design; this ensures that the system remains operational in the presence of benign failures or any resource depletion attack, and that it dynamically \emph{scales out}, or possibly \emph{scales in}, according to the requests' arrival rate. Our full-blown implementation on the Google Cloud Platform shows that deploying a \ac{VPKI} for a large-scale scenario can be cost-effective, while efficiently issuing pseudonyms for the requesters. 

\end{abstract}


\maketitle

\section{Introduction}
\label{sec:vpkiaas-demo-introduction}

\acresetall

In \ac{VC} systems, vehicles beacon \acp{CAM} periodically, at a high rate, to enable transportation safety and efficiency. \ac{V2X} communication is protected with the help of Public Key Cryptography: a set of short-lived anonymized certificates, termed \emph{pseudonyms}, are issued by a \ac{VPKI}, e.g.,~\cite{khodaei2018Secmace}, for registered vehicles. Vehicles switch from one pseudonym to a non-previously used one towards message unlinkability as pseudonyms are per se inherently unlinkable. 

With emerging large-scale multi-domain \ac{VC} environments, the efficiency of the \ac{VPKI} and, more broadly, its scalability are paramount. Deploying a \ac{VPKI} differs from a traditional \acs{PKI} in different aspects. One of the most important factors is the dimension of the \acs{PKI}, i.e., the number of registered ``users'' (vehicles) and the multiplicity of certificates per user. According to the US \ac{DoT}, a \ac{VPKI} should be able to issue pseudonyms for more that 350 million vehicles across the Nation~\cite{DOTHS812014}. Considering the average daily commute time to be 1 hour~\cite{DOTHS812014} and a pseudonym lifetime of 5 min, the \ac{VPKI} should be able to issue at least $1.5 \times 10^{12}$ pseudonyms per year\footnote{Note that this number could be even greater by considering the envisioned vehicular ecosystem, i.e., pedestrian and cyclist being part of the \acp{ITS} with a gamut of services, e.g., \acp{LBS}.}, i.e., 5 orders of magnitude more than what the largest current \acs{PKI} issues (10 million certificates per year~\cite{whyte2013security}).

Each vehicle is expected to interact with the \ac{VPKI} regularly, e.g., once or a few times per day, not only to refill its pseudonym pool, but also to fetch the latest revocation information. As shown in~\cite{khodaei2014ScalableRobustVPKI, khodaei2018Secmace}, the performance of a \ac{VPKI} system can be drastically degraded under a clogging \ac{DoS} attack: adversaries could compromise the availability of the \ac{VPKI} entities with spurious requests. The cost of unavailability\footnote{Note that the \ac{VPKI} could be unreachable for other reasons, e.g., intermittent coverage of sparsely-deployed \acp{RSU}, that are orthogonal to this investigation.} is twofold: security (degradation of road safety) and privacy. An active malicious entity could prevent other vehicles from accessing the \ac{VPKI} to fetch the latest revocation information. Moreover, signing \acp{CAM} with the private keys corresponding to expired pseudonyms, or the \ac{LTC}, is insecure and harms user privacy. Even though one can refill its pseudonym pool by relying on other anonymous authentication primitives, e.g.,~\cite{khodaei2017RHyTHM}, the performance of the safety-related applications could be degraded if the majority of vehicles leverage such schemes, i.e., causing 30\% increase in cryptographic processing overhead in order to validate \acp{CAM}~\cite{khodaei2017RHyTHM}.

In this work, we leverage and \emph{enhance} the state-of-the-art \ac{VPKI} towards a highly-available, dynamically-scalable, and fault-tolerant (highly-reliable) design to ensure that the system remains operational in the presence of benign failures or any resource depletion attack (clogging \ac{DoS}). Moreover, we show how to dynamically scale out, or possibly scale in\footnote{Scaling in/out, termed \emph{horizontal} scaling, refers to replicating a new instance of a service, while scaling up/down, termed \emph{vertical} scaling, refers to allocating/deallocating resources for an instance of a given service.}, based on the workload on the \ac{VPKI} system, so that it can comfortably handle any demanding load while being cost-effective by systematically allocating/deallocating resources.

\section{VPKI\lowercase{ as a }Service (VPKI\lowercase{aa}S)}
\label{sec:vpkiaas-demo-system-model}

We leverage the state-of-the-art \ac{VPKI} system~\cite{khodaei2018Secmace} that provides pseudonyms in an \emph{on-demand} fashion: each vehicle \emph{``decides''} when to trigger the pseudonym acquisition process based on various factors~\cite{khodaei2016evaluating}. Pseudonyms have a lifetime (a validity period), typically ranging from minutes to hours; in principle, the shorter the pseudonym lifetime ($\tau_{P}$) is, the higher the unlinkability and thus the higher the privacy protection that can be achieved.

The \ac{VPKI} consists of a set of Certification Authorities (CAs) with distinct roles: the \ac{RCA}, the highest-level authority, certifies other lower-level authorities; the \ac{LTCA} is responsible for the vehicle registration, the \acf{LTC} issuance, as well as (authorization) \emph{ticket} issuance, used for obtaining pseudonyms. The \ac{PCA} issues pseudonyms for the registered vehicles and the \acf{RA} can initiate a process to resolve and revoke all pseudonyms of a misbehaving vehicle. Vehicles can cross into other \emph{domains}~\cite{khodaei2015VTMagazine}; trust between two domains can be established with the help of an \ac{RCA}, or through cross certification between them~\cite{khodaei2015VTMagazine}. The efficiency and robustness of the \ac{VPKI} system is systematically investigated in~\cite{khodaei2016evaluating, khodaei2018Secmace} and the \ac{VPKI} can handle large workloads. A detailed protocol description can be found in~\cite{khodaei2014ScalableRobustVPKI, khodaei2018Secmace}.

\begin{figure} [!t]
	\vspace{-0.5em}
	\begin{center}
		\centering
		\includegraphics[trim=0cm 0cm 0cm 0cm, clip=true, totalheight=0.22\textheight,angle=0]{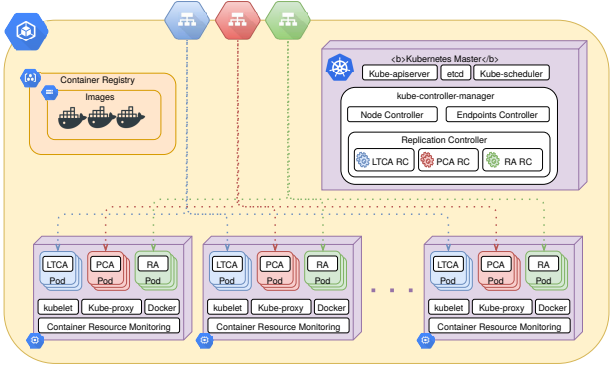}
		\vspace{-0.75em}
		\caption{A high-level \acs{VPKIaaS} architecture.}
		\label{fig:vpkiaas-demo-system-design}
		\vspace{-1.25em}
	\end{center}
\end{figure}

Towards ensuring viability as \ac{VC} systems grow, we deploy the \ac{VPKI} on the \ac{GCP} (cloud.google.com), and evaluate the availability, reliability, and dynamic scalability of our scheme under various circumstances. Fig.~\ref{fig:vpkiaas-demo-system-design} illustrates a high-level abstraction of the \acs{VPKIaaS} architecture on a managed Kubernetes cluster (kubernetes.io) on \ac{GCP}.\footnote{The \ac{RCA} entity is assumed to be off-line, thus not included in the architecture.} A set of Pods will be created for each micro-service, e.g., \ac{LTCA} or \ac{PCA}, from their corresponding container images, facilitating their horizontal scalability. When the rate of pseudonym requests increases, the Kubernetes master, shown on the top, schedules new Pods or kills a running Pod in case of benign failures, e.g., system faults or crashes, or resource depletion attacks, e.g., a \ac{DoS} attack. The Pods could be scaled out to the number set in the deployment configuration, or the amount of available resources enabled by Kubernetes nodes. 

Each Pod publishes two types of metrics: \emph{load} and \emph{health}. The load metric values are generated by a resource monitoring service, which facilitates horizontal scaling of a micro-service, i.e., upon reaching a threshold of a defined load, replication controller replicates a new instance of the micro-service to ensure a desired \ac{SLA}. Health metric ensures correct operation of a micro-service by persistently monitoring its status: a faulty Pod is killed and a new one will be created. In our system, we defined CPU and memory usage as the load metric. In order to monitor the health condition of a micro-service, dummy requests (tickets for \ac{LTCA} micro-services and pseudonyms for \ac{PCA} micro-services) are queried (locally by each micro-service).

\textbf{Note:} Multiple replicas of a micro-service interact with the same database to accomplish their operations, e.g., all replicas of \acp{PCA} interact with a single database to validate an authorization ticket and store information corresponding to the issued pseudonyms. This could be a bottleneck in our architecture, possibly a single point of failure; how to mitigate this would be part of our future investigation. Moreover, the information corresponding to the issued pseudonyms is stored asynchronously, i.e., a \ac{PCA} micro-service delivers the pseudonym response without confirmation of its successful storage in the database. If there are multiple replicas of a micro-service, e.g., a \ac{PCA}, a \emph{``malicious''} vehicle, repeatedly requesting to obtain pseudonyms, might be provided with more than a set of pseudonyms per ticket.\footnote{This vulnerability is also relevant to the ticket acquisition process.} As future work, we plan to utilize alternative storage solutions, e.g., \emph{NoSQL} databases, and apply zonal resource synchronization to prevent issuing multiple pseudonyms per ticket, thus fully mitigating the vulnerability.

\section{Demonstration}
\label{sec:vpkiaas-demo-demonstration}

In this work, we demonstrate three scenarios: pseudonym acquisition using Nexcom vehicular \acp{OBU} (S1), Pseudonym acquisition for a large-scale urban vehicular mobility dataset (S2), and the performance of the \acs{VPKIaaS} system, notably its \emph{high-availability}, \emph{robustness}, \emph{reliability}, and \emph{dynamic-scalability} (S3). For the first two scenarios, our metric is the end-to-end pseudonym acquisition latency, measured at the \ac{OBU} side. For the last scenario, we aim at demonstrating the performance of our \acs{VPKIaaS} system by emulating a large volume of workload. For each scenario, the authors (presenters) will explain the underlying concepts behind different components of our scheme along with the achieved results. 

\textbf{Experimental setup:} We created and pushed Docker images for \ac{LTCA}, \ac{PCA}, \ac{RA}, and MySQL to the Google Container Registry (cloud.google.com/container-registry/). Isolated namespaces and deployment configuration files are defined before Google Kubernetes Engine v1.9.6 (cloud.google.com/kubernetes-engine) cluster runs the workload. We configured a cluster of three \acp{VM}, each with eight vCPUs and 10GB of memory. To emulate a large volume of workload, we created another Kubernetes cluster of four \acp{VM} (in a different data center), each with 10 vCPUs and 10GB of memory. Our full-blown implementation is in C++ and we use FastCGI~\cite{heinlein1998fastcgi} to interface Apache web-server. We use XML-RPC (xmlrpc-c.sourceforge.net) to execute a remote procedure call on the cloud. Our \acs{VPKIaaS} interface is language-neutral and platform-neutral, as we use Protocol Buffers (developers.google .com/protocol-buffers) for serializing and de-serializing structured data. For the cryptographic protocols and primitives (\ac{ECDSA} and \acs{TLS}), we use OpenSSL with \ac{ECDSA}-256 public/private key pairs according to the ETSI (TR-102-638) and IEEE 1609.2 standards; other algorithms and key sizes are compatible in our implementation. 

\subsection{S1: Pseudonym Acquisition by an \ac{OBU}}
\label{sec:vpkiaas-demo-scenario1}

Fig.~\ref{fig:vpkiaas-demo-end-to-end-latency-to-obtain-pseudonyms-lust-dataset}.a shows two Nexcom vehicular \acp{OBU} (Dual-core 1.66 GHz, 1GB memory) from PRESERVE project (www.preserve-project.eu), which support IEEE 802.11p. In this scenario, we consider one of them to be an \ac{RSU}, connected to the \ac{VPKI} via Ethernet. The second \ac{OBU} requests pseudonyms from the \ac{VPKI} via the ``\ac{RSU}'' and the communication is over IEEE 802.11p.

\begin{figure} [!t]
	\vspace{-0.5em}
	\begin{center}
		\centering
		\subfloat[]{
			\hspace{-1em} \includegraphics[trim=0cm 0.2cm 0.75cm 1cm, clip=true, width=0.25\textwidth,height=0.25\textheight,angle=0,keepaspectratio]{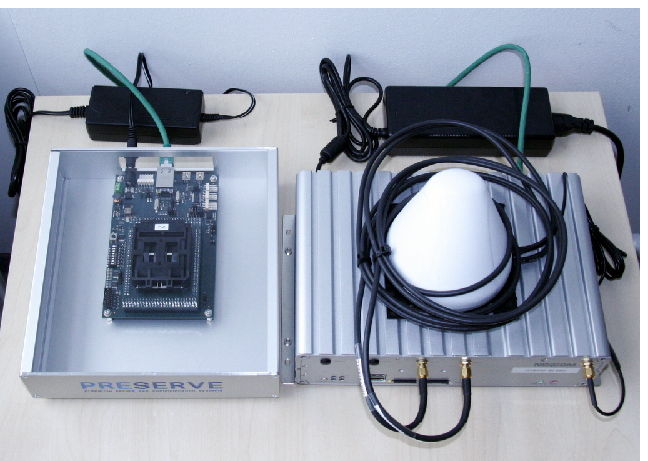}} 
		\hfill
		\subfloat[] {
			\hspace{-0.75em} \includegraphics[trim=0cm 0.2cm 2.75cm 1cm, clip=true, width=0.23\textwidth,height=0.23\textheight,angle=0,keepaspectratio]{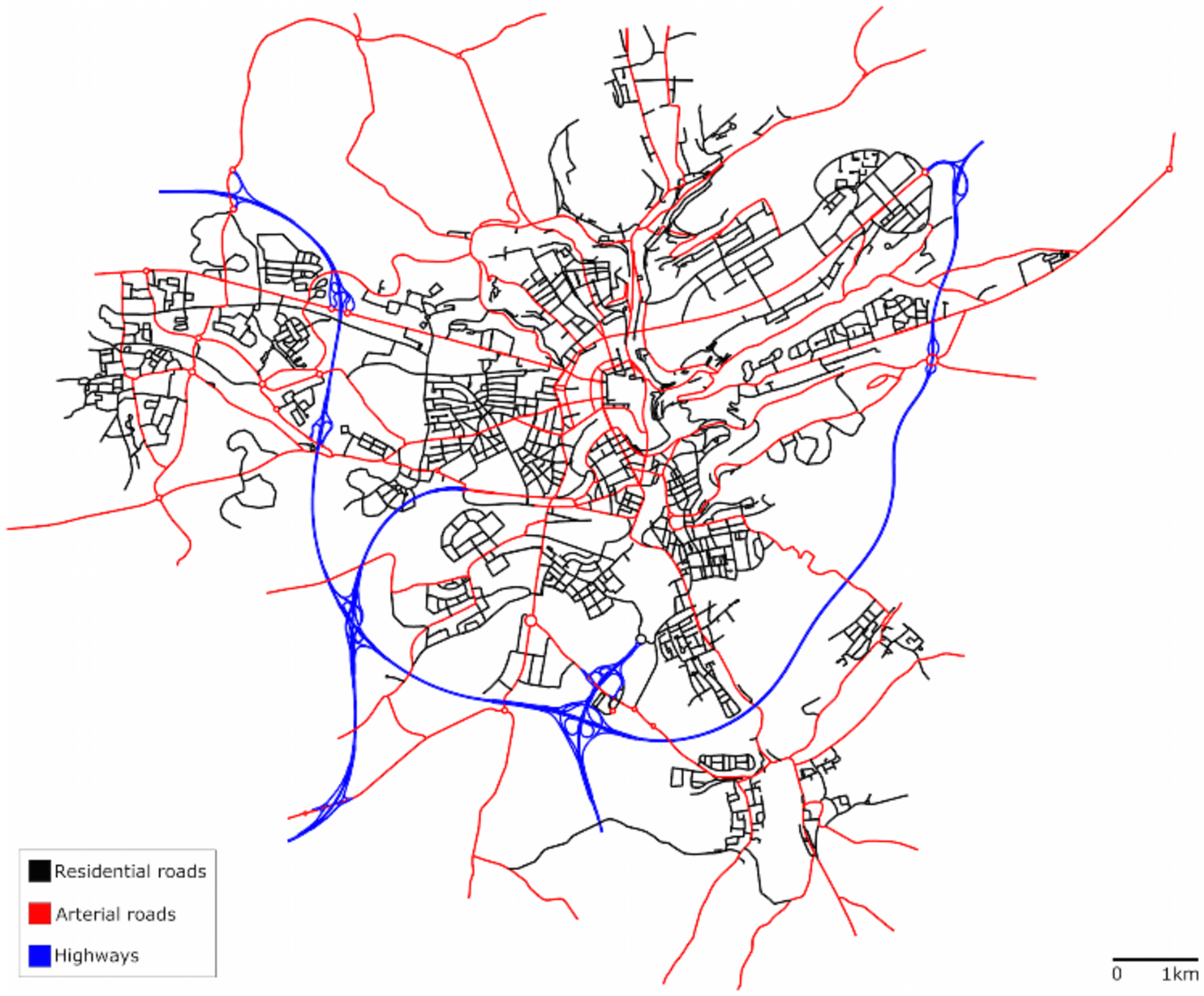}}
		\vspace{-1em}
		\caption{(a) Nexcom boxes from the PRESERVE project, used in S1. (b) \acs{LuST} Topology~\cite{codeca2015lust}, used in S2.}
		\label{fig:vpkiaas-demo-nexcom-boxes-lust-scenario-topology}
	\end{center}
	\vspace{-1.5em}
\end{figure}

\subsection{S2: Large-scale Pseudonym Acquisition}
\label{sec:vpkiaas-demo-scenario2}

\begin{figure} [!t]
	\vspace{-0em}
	\begin{center}
		\centering
		\subfloat[]{
			\hspace{-1.25em} \includegraphics[trim=0cm 0.2cm 0.75cm 1cm, clip=true, totalheight=0.15\textheight,angle=0,keepaspectratio]{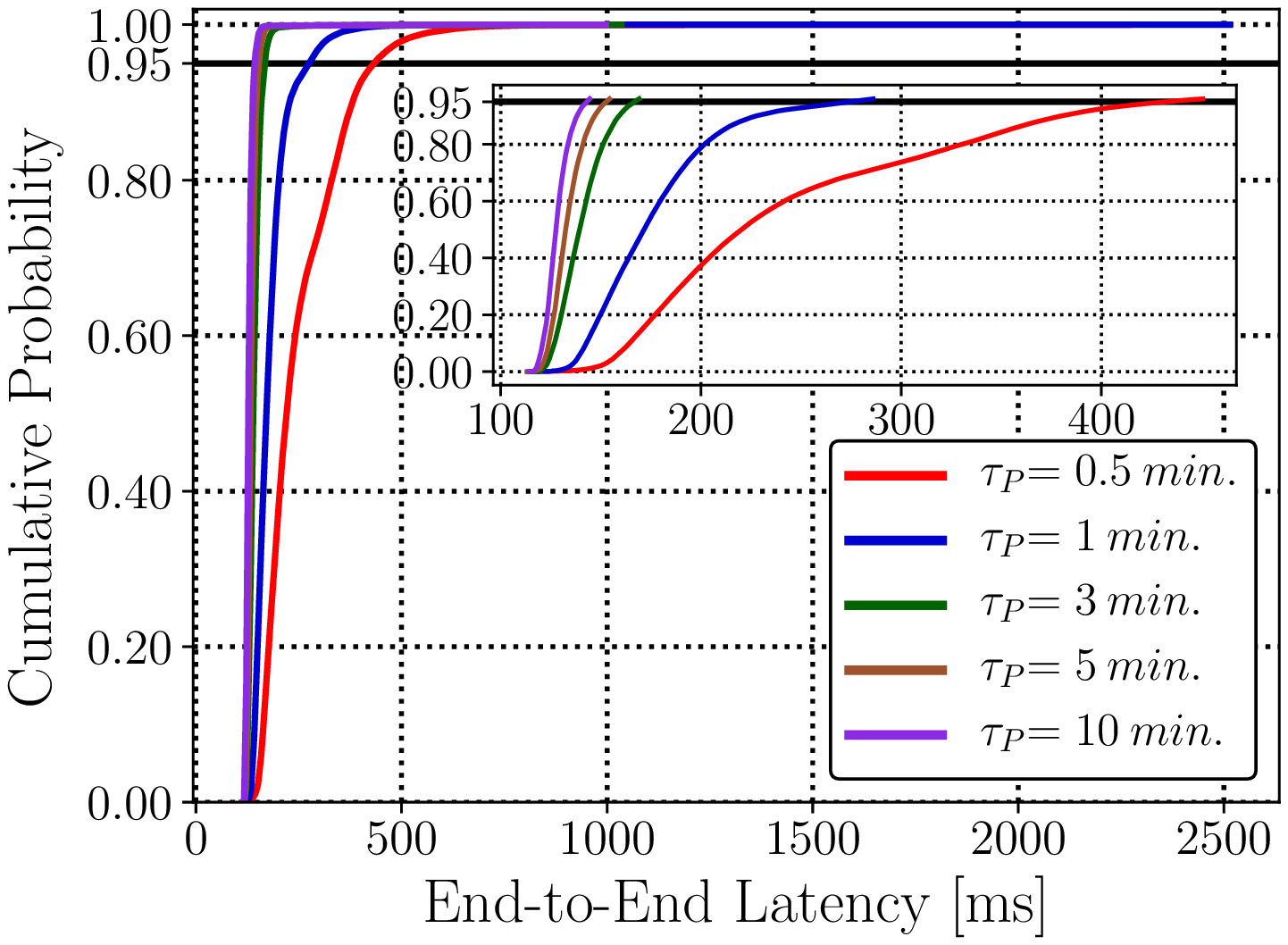}}
		\subfloat[] {
			\hspace{-0.75em} \includegraphics[trim=0cm 0.2cm 0.75cm 1cm, clip=true, totalheight=0.15\textheight,angle=0,keepaspectratio]{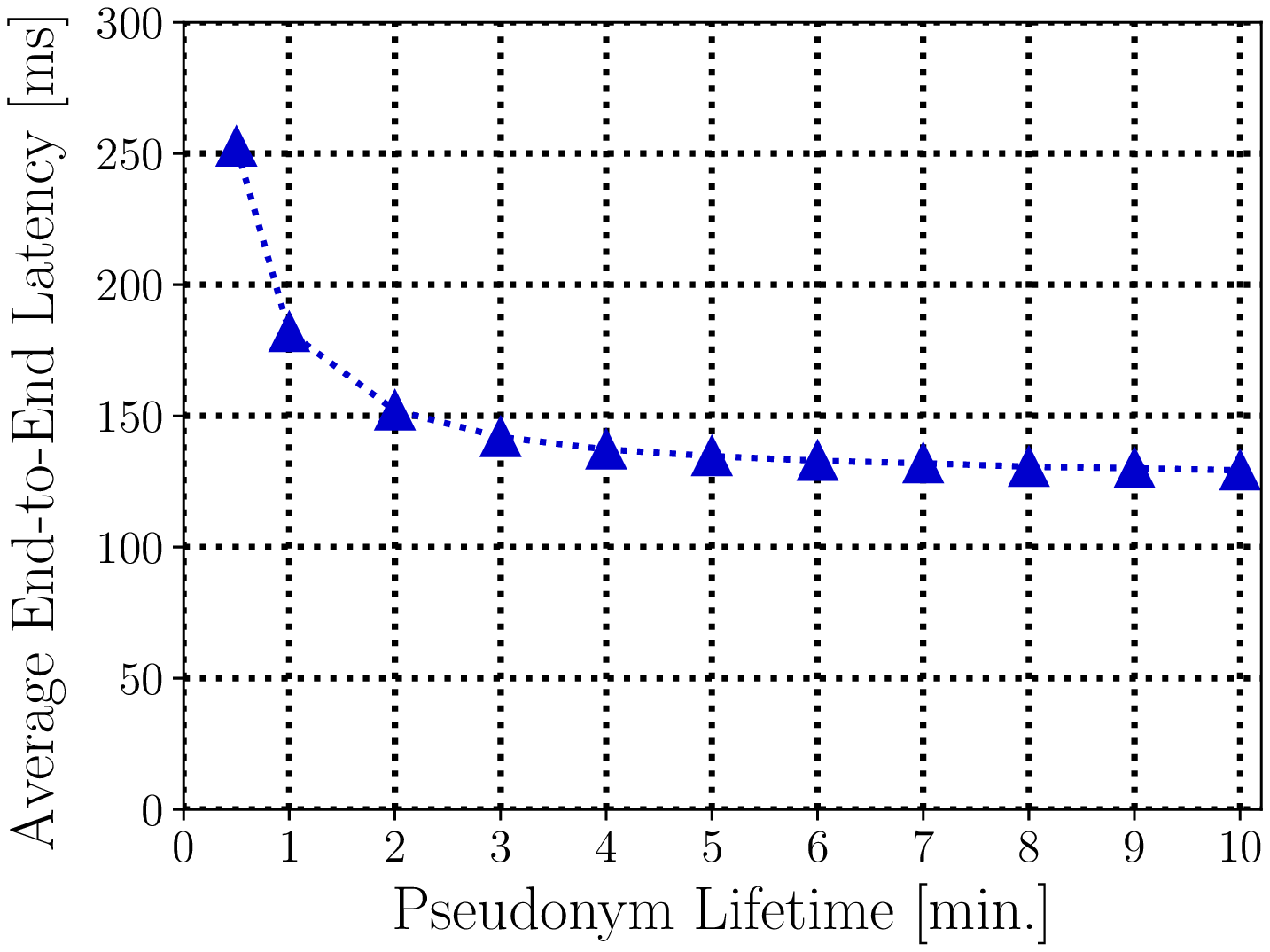}}
		\vspace{-1.25em}
		\caption{(a) End-to-end latency for pseudonym acquisition. (b) Average end-to-end latency.}
		\label{fig:vpkiaas-demo-end-to-end-latency-to-obtain-pseudonyms-lust-dataset}
	\end{center}
	\vspace{-0.5em}
\end{figure}

Fig.~\ref{fig:vpkiaas-demo-end-to-end-latency-to-obtain-pseudonyms-lust-dataset}.b shows the \acs{LuST}~\cite{codeca2015lust} scenario topology, a full-day realistic mobility pattern in the city of Luxembourg. We use OMNET++ (omnetpp.org) and the Veins framework to simulate this large-scale scenario using SUMO. In our simulation, we placed 100 \acp{RSU} in a region (50KM $\times$ 50KM). Each vehicle requests pseudonyms for its actual trip duration and \ac{V2I} communication is IEEE 802.11p.

Fig.~\ref{fig:vpkiaas-demo-end-to-end-latency-to-obtain-pseudonyms-lust-dataset}.a illustrates the CDF of the actual end-to-end latencies for obtaining pseudonyms with different pseudonyms lifetimes ($\tau_{P}$) during the rush hours (7-9 am and 5-7 pm). For example, with $\tau_{P}=1$ minute, 95\% of the vehicles are served within less than 286 ms. Fig.~\ref{fig:vpkiaas-demo-end-to-end-latency-to-obtain-pseudonyms-lust-dataset}.b shows the average end-to-end latency with different pseudonyms lifetimes. Obviously, the shorter the $\tau_{P}$, the higher the workload on the \ac{VPKI}, thus the higher the end-to-end latency. The results confirm that our scheme is efficient and scalable: the pseudonym acquisition process incurs low latency and it efficiently issues pseudonyms for the requesters.

\subsection{S3: \acs{VPKIaaS} Performance}
\label{sec:vpkiaas-demo-scenario3}

\begin{figure} [!t]
	\vspace{-0.25em}
	\begin{center}
		\centering
		\includegraphics[trim=0.15cm 0cm 0cm 1cm, clip=true, width=0.43\textwidth,height=0.43\textheight,keepaspectratio]{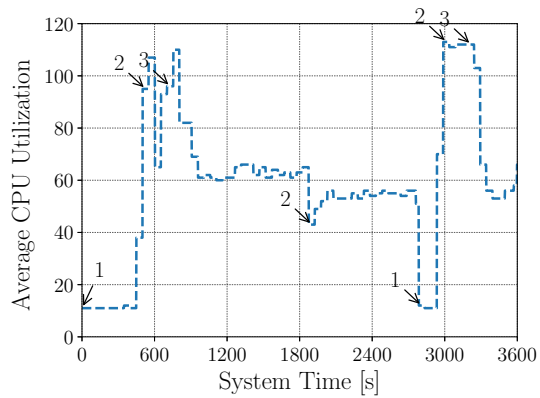}
		\vspace{-1.25em}
		\caption{Dynamic scalability of the \acs{VPKIaaS}.}
		\label{fig:vpkiaas-demo-performance-evaluation}
	\end{center}
	\vspace{-1.25em}
\end{figure}

In this scenario, we aim at demonstrating the performance of our \ac{VPKI}, notably its reliability and dynamic scalability. To emulate a large volume of workload, we generated synthetic workload with up to 14 containers with 1-4 vCPUs and 1-4 GB of memory. Each container generates 80,000 requests in the span of one hour leveraging 16 threads. One pseudonym request encapsulates 100 \acp{CSR} according to the standard (ETSI TR-102-638 and IEEE 1609.2). Fig.~\ref{fig:vpkiaas-demo-performance-evaluation} shows how our \ac{VPKI} system dynamically scales out/in according to the rate of pseudonym requests. The numbers next to the arrows show the number of \ac{PCA} Pods at a specific system time. 

We achieve a 5-fold improvement over prior work~\cite{cincilla2016vehicular}: the processing delay to issue a pseudonym for~\cite{cincilla2016vehicular} is 20 ms, while it is approx. 4 ms in our system. Moreover, unlike the \ac{VPKI} system in~\cite{cincilla2016vehicular}, our implementation supports dynamic scalability, i.e., the \ac{VPKI} scales in/out based on the arrival rate of pseudonym requests.

\section{Conclusion}
\label{sec:vpkiaas-demo-conclusions}

Paving the way for the deployment of a secure and privacy-preserving \ac{VC} system relies on deploying a special-purpose \ac{VPKI}. However, its success requires extensive experimental evaluation, to ensure viability (in terms of performance and cost). We leverage the state-of-the-art \ac{VPKI} and show its availability, resiliency, and scalability towards a cost-effective \ac{VPKI} deployment.

\section*{Acknowledgement}
\label{sec:vpkiaas-demo-acknowledgement}

This work has been partially supported by the Swedish Foundation for Strategic Research (SSF).

\bibliographystyle{ACM-Reference-Format}
\bibliography{references} 


\begin{thebibliography}{00}


\ifx \showCODEN    \undefined \def \showCODEN     #1{\unskip}     \fi
\ifx \showDOI      \undefined \def \showDOI       #1{{\tt DOI:}\penalty0{#1}\ }
  \fi
\ifx \showISBNx    \undefined \def \showISBNx     #1{\unskip}     \fi
\ifx \showISBNxiii \undefined \def \showISBNxiii  #1{\unskip}     \fi
\ifx \showISSN     \undefined \def \showISSN      #1{\unskip}     \fi
\ifx \showLCCN     \undefined \def \showLCCN      #1{\unskip}     \fi
\ifx \shownote     \undefined \def \shownote      #1{#1}          \fi
\ifx \showarticletitle \undefined \def \showarticletitle #1{#1}   \fi
\ifx \showURL      \undefined \def \showURL       #1{#1}          \fi
\providecommand\bibfield[2]{#2}
\providecommand\bibinfo[2]{#2}
\providecommand\natexlab[1]{#1}
\providecommand\showeprint[2][]{arXiv:#2}

\bibitem[\protect\citeauthoryear{??}{DOT}{2014}]%
        {DOTHS812014}
 \bibinfo{year}{2014}\natexlab{}.
\newblock \bibinfo{title}{{V2V} {C}ommunications: {R}eadiness of {V2V}
  {T}echnology for {A}pplication}.
\newblock   (\bibinfo{date}{Aug.} \bibinfo{year}{2014}).
\newblock
\newblock
\shownote{{N}ational {H}ighway {T}raffic {S}afety {A}dministration, {DOT HS}
  812 014.}


\bibitem[\protect\citeauthoryear{Cincilla and et~al}{Cincilla and
  et~al}{2016}]%
        {cincilla2016vehicular}
\bibfield{author}{\bibinfo{person}{P. Cincilla} {and} \bibinfo{person}{et al}.}
  \bibinfo{year}{2016}\natexlab{}.
\newblock \showarticletitle{{Vehicular PKI Scalability-Consistency Trade-Offs
  in Large Scale Distributed Scenarios}}. In \bibinfo{booktitle}{{\em IEEE
  VNC}}. \bibinfo{address}{Columbus, Ohio, USA}.
\newblock


\bibitem[\protect\citeauthoryear{Codeca and et~al}{Codeca and et~al}{2015}]%
        {codeca2015lust}
\bibfield{author}{\bibinfo{person}{L. Codeca} {and} \bibinfo{person}{et al}.}
  \bibinfo{year}{2015}\natexlab{}.
\newblock \showarticletitle{{L}uxembourg {S}UMO {T}raffic ({LuST}) {S}cenario:
  24 {H}ours of {M}obility for {V}ehicular {N}etworking {R}esearch}. In
  \bibinfo{booktitle}{{\em IEEE VNC}}. \bibinfo{address}{Kyoto, Japan}.
\newblock


\bibitem[\protect\citeauthoryear{Heinlein}{Heinlein}{1998}]%
        {heinlein1998fastcgi}
\bibfield{author}{\bibinfo{person}{Paul Heinlein}.}
  \bibinfo{year}{1998}\natexlab{}.
\newblock \showarticletitle{{F}ast{CGI}}.
\newblock \bibinfo{journal}{{\em Linux journal\/}} \bibinfo{volume}{1998},
  \bibinfo{number}{55es} (\bibinfo{year}{1998}), \bibinfo{pages}{1}.
\newblock


\bibitem[\protect\citeauthoryear{Khodaei and et~al}{Khodaei and et~al}{2014}]%
        {khodaei2014ScalableRobustVPKI}
\bibfield{author}{\bibinfo{person}{M. Khodaei} {and} \bibinfo{person}{et al}.}
  \bibinfo{year}{2014}\natexlab{}.
\newblock \showarticletitle{{T}owards {d}eploying a {s}calable \& {r}obust
  {v}ehicular {i}dentity and {c}redential {m}anagement {i}nfrastructure}. In
  \bibinfo{booktitle}{{\em VNC}}. \bibinfo{address}{Paderborn,$\;\;$Germany}.
\newblock


\bibitem[\protect\citeauthoryear{Khodaei and et~al}{Khodaei and et~al}{2016}]%
        {khodaei2016evaluating}
\bibfield{author}{\bibinfo{person}{M. Khodaei} {and} \bibinfo{person}{et al}.}
  \bibinfo{year}{2016}\natexlab{}.
\newblock \showarticletitle{{E}valuating {O}n-demand {P}seudonym {A}cquisition
  {P}olicies in {V}ehicular {C}ommunication {S}ystems}. In
  \bibinfo{booktitle}{{\em IoV/VoI}}. \bibinfo{address}{Paderborn, Germany}.
\newblock


\bibitem[\protect\citeauthoryear{Khodaei and et~al.}{Khodaei and
  et~al.}{2018}]%
        {khodaei2018Secmace}
\bibfield{author}{\bibinfo{person}{M. Khodaei} {and} \bibinfo{person}{et al.}}
  \bibinfo{year}{2018}\natexlab{}.
\newblock \showarticletitle{{SECMACE}: {S}calable and {R}obust {I}dentity and
  {C}redential {M}anagement {I}nfrastructure in {V}ehicular {C}ommunication
  {S}ystems}.
\newblock \bibinfo{journal}{{\em IEEE Transactions on Intelligent
  Transportation Systems\/}} \bibinfo{volume}{19}, \bibinfo{number}{5}
  (\bibinfo{date}{May} \bibinfo{year}{2018}), \bibinfo{pages}{1430--1444}.
\newblock


\bibitem[\protect\citeauthoryear{Khodaei, Messing, and Papadimitratos}{Khodaei
  et~al\mbox{.}}{2017}]%
        {khodaei2017RHyTHM}
\bibfield{author}{\bibinfo{person}{M. Khodaei}, \bibinfo{person}{A. Messing},
  {and} \bibinfo{person}{P. Papadimitratos}.} \bibinfo{year}{2017}\natexlab{}.
\newblock \showarticletitle{{RHyTHM}: {A} {R}andomized {H}ybrid {S}cheme {T}o
  {H}ide in the {M}obile {C}rowd}. In \bibinfo{booktitle}{{\em IEEE VNC}}.
  \bibinfo{address}{Torino, Italy}.
\newblock


\bibitem[\protect\citeauthoryear{Khodaei and Papadimitratos}{Khodaei and
  Papadimitratos}{2015}]%
        {khodaei2015VTMagazine}
\bibfield{author}{\bibinfo{person}{M. Khodaei} {and} \bibinfo{person}{P.
  Papadimitratos}.} \bibinfo{year}{2015}\natexlab{}.
\newblock \showarticletitle{{T}he {K}ey to {I}ntelligent {T}ransportation:
  {I}dentity and {C}redential {M}anagement in {V}ehicular {C}ommunication
  {S}ystems}.
\newblock \bibinfo{journal}{{\em IEEE VT Magazine\/}} \bibinfo{volume}{10},
  \bibinfo{number}{4} (\bibinfo{date}{Dec.} \bibinfo{year}{2015}),
  \bibinfo{pages}{63--69}.
\newblock


\bibitem[\protect\citeauthoryear{Whyte, Weimerskirch, Kumar, and Hehn}{Whyte
  et~al\mbox{.}}{2013}]%
        {whyte2013security}
\bibfield{author}{\bibinfo{person}{W. Whyte}, \bibinfo{person}{A Weimerskirch},
  \bibinfo{person}{V. Kumar}, {and} \bibinfo{person}{T. Hehn}.}
  \bibinfo{year}{2013}\natexlab{}.
\newblock \showarticletitle{{A} {S}ecurity {C}redential {M}anagement {S}ystem
  for {V2V} {C}ommunications}. In \bibinfo{booktitle}{{\em IEEE VNC}}.
  \bibinfo{address}{Boston, MA}.
\newblock


\end{thebibliography}

\end{document}